\documentclass[%
reprint,
 amsmath,amssymb,
 aps, 
 physrev,
]{revtex4-2}
\usepackage{graphicx}
\usepackage{dcolumn}
\usepackage{bm}
\usepackage{hyperref}
\usepackage[mathlines]{lineno}

\usepackage{siunitx}
\usepackage{physics}
\usepackage{amsmath}
\usepackage{isotope}
\usepackage{stmaryrd}
\usepackage{float}

\newcommand{\qunaught}{\ensuremath{\ket{\varnothing}}}

\begin{document}

\preprint{APS/123-QED}

\title{\textbf{Error Correction of Beamsplitter-Generated Entangled GKP States}}

\author{Moritz Fontboté-Schmidt}
\altaffiliation{These authors contributed equally.}
\email{Contact author: moritzfo@phys.ethz.ch}
\author{Jeremy Metzner}
\altaffiliation{These authors contributed equally.}
\author{Florence Berterottière}
\altaffiliation{These authors contributed equally.}

\author{Ivan Rojkov}
\author{Alexander Ferk}
\author{Martin Stadler}
\author{Bahadir Dönmez}

\author{Ralf Berner}
\altaffiliation{Present address: 5. Physikalisches Institut, Universität Stuttgart, Pfaﬀenwaldring 57, 70569 Stuttgart, Germany.}

\author{Stephan Welte}
\altaffiliation{Present address: 5. Physikalisches Institut, Universität Stuttgart, Pfaﬀenwaldring 57, 70569 Stuttgart, Germany.}
\author{Daniel Kienzler}
\author{Jonathan P. Home}%

\affiliation{%
Institute for Quantum Electronics, ETH Zürich, Otto-Stern-Weg 1, 8093 Zürich, Switzerland and Quantum Center, ETH Zürich, 8093 Zürich, Switzerland}%

\maketitle

\noindent \textbf{ 
To be useful, quantum computers will be required to successfully correct errors occurring at the hardware level. Bosonic codes provide a hardware-efficient option for error correction, but fault-tolerance further requires that the available gate interactions be compatible with the code. A promising bosonic code is the Gottesman-Kitaev-Preskill (GKP) code~\cite{gottesman_encoding_2001} for which a linear beamsplitter-like coupling between two bosonic modes is fault-tolerant \cite{gottesman_encoding_2001,glancy_error_2006, menicucci_fault-tolerant_2014,walshe_continuous-variable_2020, shaw_logical_2024}, which makes this a key primitive for building larger systems. Here, using two motional modes of a trapped ion, we demonstrate the generation of entangled states of GKP qubits by interfering two qunaught states\cite{walshe_continuous-variable_2020}, which have a grid structure but carry no logical information, on a beamsplitter. We generate all four Bell states with an average fidelity of 69\%, and subsequently demonstrate an extension of the entangled state lifetime through the use of quantum error correction. These results complete the set of Gaussian operations required for quantum computing with GKP codes \cite{menicucci_fault-tolerant_2014} and enable explorations of multi-mode bosonic encodings~\cite{royer_encoding_2022} as well as fundamental tests of information channels \cite{wang_passive_2025}. }

\begin{figure*}[t]
    \centering
    \includegraphics[width=1\linewidth]{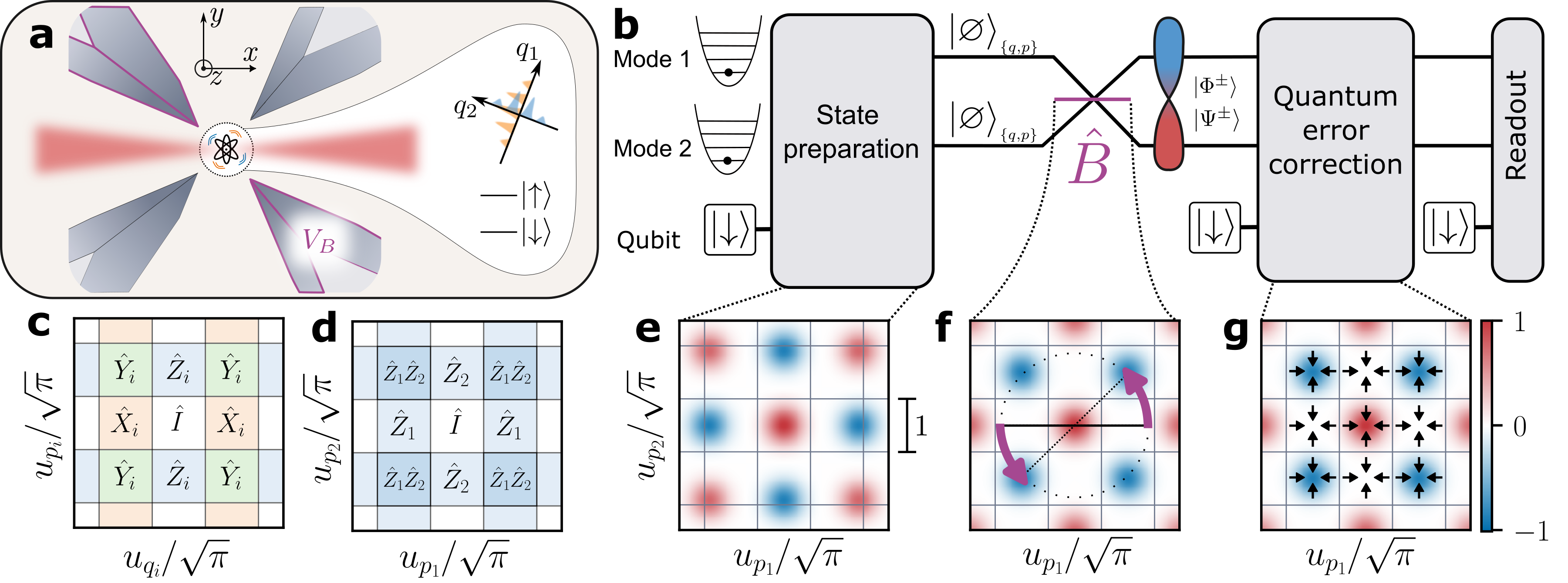}
    \caption{\textbf{Bell state protocol.} \textbf{a} Information is stored in two orthogonal mechanical modes of a single $\isotope[40]{Ca}^+$ ion, trapped at the center of a Paul trap (grey). These are coupled by modulating the trap curvature along an axis with a projection on both modes by applying voltages to selected trap electrodes (purple). A bichromatic laser beam (red) is used for coupling the ancilla qubit to the oscillator for preparation, control and readout of the GKP states. \textbf{b} Bell states of logical GKP qubits are prepared by using an interaction $\hat{B}$ analogous to an optical 50/50 beamsplitter to interfere lower dimensional qunaught states that are initially prepared in the two mechanical modes. \textbf{c} Displacements by $(u_q, u_p)$ in phase space are divided into square regions associated with the nearest logical operator expectation values of the GKP code. \textbf{d} In the correlated $u_{p_1}, u_{p_2}$ phase space, some of the elements correspond to correlated logical observables. \textbf{e} Theoretical two-mode characteristic functions of the qunaught product state in the $u_{p_1},u_{p_2}$ plane, which the beamsplitter operation rotates into \textbf{f} a Bell state. \textbf{g} Error correction (black arrows) prevents decoherence.   }
    \label{fig 1: overview}
\end{figure*}

Useful quantum computers will require error-correction to work reliably on a relevant scale~\cite{preskill_quantum_2018}. This is expected to require considerable resource overhead, even while using high-quality operations \cite{gidney_how_2025, reiher_elucidating_2017}. This motivates the search for physically compact codes that mitigate these requirements. Bosonic codes address this challenge by exploiting the infinite-dimensional Hilbert space of a single quantum harmonic oscillator to achieve redundancy without multiplying the number of physical systems~\cite{albert_bosonic_2025}. To work fault-tolerantly, such codes must be capable of correcting errors introduced by the quantum gates which operate on them~\cite{gottesman_quantum_2005}. The Gottesman-Kitaev-Preskill error-correction code is a promising candidate, because it is capable of correcting all low-order errors which occur, with relatively simple physical operations. The most basic realization of the GKP code imposes a two-dimensional discrete translational symmetry forming a lattice in the position-momentum phase space, and protects information from small displacements which are less than a quarter of the lattice constant of the code \cite{gottesman_encoding_2001, fluhmann_encoding_2019, royer_stabilization_2020, de_neeve_error_2022, sivak_real-time_2023, brock_quantum_2025, matsos_robust_2024, matsos_universal_2025}. 

Given these features, GKP codes have attracted significant interest, with theoretical work outlining a path to large-scale quantum computers by combining these states with networks of beamsplitters and other Gaussian operations~\cite{menicucci_fault-tolerant_2014, walshe_continuous-variable_2020, walshe_linear-optical_2025}. This path remains available even when considering that experimental implementations of GKP codes cannot be fully translationally invariant, due to the finite energy of the states involved. The first demonstrations of GKP qubits were performed with trapped-ions and superconducting circuits~\cite{fluhmann_encoding_2019,campagne-ibarcq_quantum_2020}, including quantum error correction~\cite{de_neeve_error_2022,campagne-ibarcq_quantum_2020}, which has now exceeded the break-even point at which the error-corrected qubit out-performs its physical constituents~\cite{sivak_real-time_2023,brock_quantum_2025}. States with a similar grid-like structure have also recently been implemented in optical platforms~\cite{larsen_integrated_2025}.  However, a direct interaction between GKP states has not yet been shown, with two-qubit gates thus far performed in an inherently non-fault-tolerant manner, using a bare (un-encoded) qubit as an intermediate information transfer element~\cite{matsos_universal_2025}. To perform gates in a manner that is compatible with fault-tolerance, an interaction that produces correctable errors is required. One such example is the direct beamsplitter interaction, which resonantly exchanges energy between the two modes in a manner that can preserve the code structure. The beamsplitter is a key primitive for bosonic quantum computing: combined with single-mode squeezing and displacements, it generates the full set of Gaussian operations on the two-mode phase space \cite{arvind_real_1995, braunstein_squeezing_2005, baragiola_all-gaussian_2019}. 

In this article, we demonstrate the use of a beamsplitter between two trapped-ion mechanical oscillators to generate entanglement between GKP qubits. By applying a beamsplitter to appropriate separable combinations of GKP qunaught states which encode no logical information, but have the appropriate structure when rescaled by the beamsplitter (Fig.~\ref{fig 1: overview} a,b),  we prepare all four Bell states with an average fidelity of 0.69(1). We show that subsequent rounds of error correction on the oscillator modes extends the logical lifetime of the entangled state by a factor of 2.0(2) beyond the bare coherence time of the Bell state. The elements used here are key components for fault-tolerant multi-qubit gates on GKP qubits\cite{rojkov_two-qubit_2024}, as well as laying the basis for investigating fundamental protocols in quantum transduction and sensing~\cite{zheng_gaussian_2023,noh_encoding_2020,wang_passive_2025}.

The protocol we implement involves preparing qunaught states $\qunaught$ in two bosonic modes, followed by the application of a beamsplitter operation between these modes. The GKP states that we generate are eigenstates of the displacement operators $\hat S_{q_i} = e^{il \sqrt{d} \hat p_i}, \hat S_{p_i} = e^{i l\sqrt{d}\hat q_i}$, where the quadratures of the modes $i\in \{1,2\}$ satisfy $[\hat q_i, \hat p_j] = i\delta_{ij}$, $l=\sqrt{2\pi}$ and $d$ is the dimension of the code. These states have a periodic grid-like structure in position-momentum phase space with a lattice spacing $l \sqrt{d}$. For $d=2$, the stabilized subspace is that of a qubit, with logical operators $\hat Z_i = e^{i\sqrt \pi \hat q_i}, \hat X_i = e^{-i\sqrt \pi \hat p_i}$ commuting in a way that forms a qubit algebra $\hat X_i \hat Z_j =  \hat Z_i \hat X_j (-1)^{ \delta_{ij}}$. For $d = 1$, four qunaught states corresponding to $\pm 1, \pm 1$ eigenstates of $\hat S_{q_i}, \hat S_{p_i}$ can be formed. In what follows, we write these states as $\qunaught$, $\qunaught_q$, $\qunaught_p$ and $\qunaught _{qp}$, where the subscript corresponds to the operator with a negative eigenvalue. Representing these states is most natural using the two-mode characteristic function $\chi(\vec u_1,\vec u_2) = \text{Tr}\left[e^{i(u_{p_1} \hat q_1 - u_{q_1}\hat p_1)}e^{i(u_{p_2} \hat q_2 - u_{q_2}\hat p_{2})}\hat \rho_{12} \right]$, where expectation values of logical and stabilizer operators are related to points in phase space corresponding to displacement amplitudes $\vec u_i = (u_{q_i}, u_{p_i})^\top$, with $u_{q_i}$ and $u_{p_i}$ being displacements along the quadratures $q_i$ and $p_i$ respectively (Fig. \ref{fig 1: overview} c,d). The 50/50 beamsplitter $\hat{B} = e^{-i\pi/4 (\hat q_1 \hat p_2 - \hat p_1 \hat q_2)}$ performs a $\pi/4$ rotation in the two-mode phase space, transforming $u'_{p_1} \rightarrow \tfrac{1}{\sqrt 2}(u_{p_1}  - u_{p_2})$, $u'_{p_2} \rightarrow \tfrac{1}{\sqrt 2}(u_{p_1}  + u_{p_2})$ and similarly for the other coordinates. Since the qunaught lattice differs from that of the GKP qubit code by a factor $\sqrt 2$ (Fig. \ref{fig 1: overview} e), this rotation matches the grid to the qubit lattice spacing and splits it into even and odd sublattices (Fig. \ref{fig 1: overview} f), transforming the separable input states into a GKP qubit Bell state~\cite{walshe_continuous-variable_2020,zheng_gaussian_2023}. The four input combinations $\qunaught \otimes \qunaught$, $\qunaught_q \otimes \qunaught_q$, $\qunaught_p \otimes \qunaught_p$ and $\qunaught _{qp} \otimes \qunaught_{qp}$ yield the four Bell states (see Methods). The output state exists in a two-mode Hilbert space where the structure of the correlations reflects the entanglement: Along the principal coordinates $u_{q_1}, u_{q_2}, u_{p_1}, u_{p_2}$ only stabilizer information is present, while logical correlations such as $\hat Z_1 \hat Z_2 = e^{i\sqrt \pi(\hat q_1 + \hat q_2)}$ appear along the correlated coordinates $u_{q_1} \pm u_{q_2}$. These correlations encode the Bell state (Fig. \ref{fig 1: overview} d,f). 

\begin{figure*}[t]
    \centering
    \includegraphics[width=1.0\linewidth]{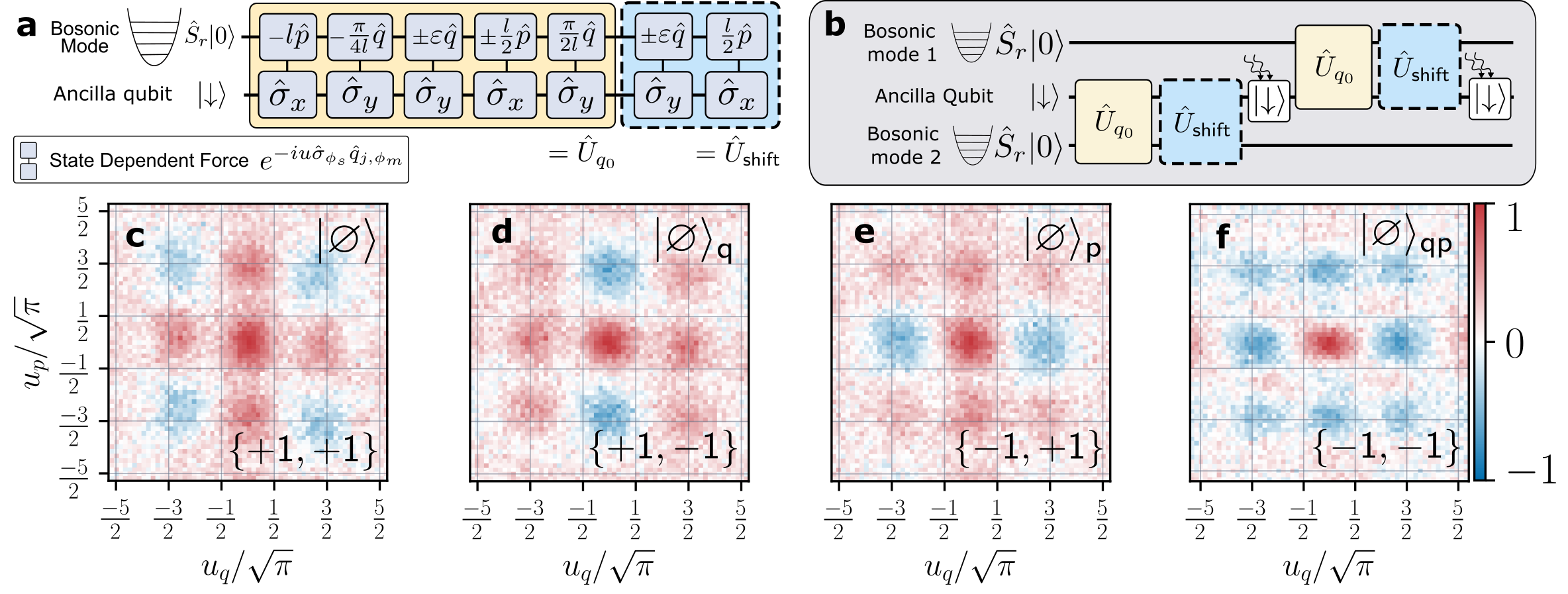}
    \caption{\textbf{Generation of different qunaught states.} \textbf{a} Experimental sequence to generate qunaught states in a single mode of motion. The sequence of SDFs generates a grid-like structure in phase space and enables control of the envelope of the state while leaving the ancilla qubit and the motion disentangled. Optional displacements at the end of the sequence enable the ability to prepare any qunaught state $\qunaught_{\{q,p\}}$ (dashed boxes) \textbf{b} Preparation of qunaught states in both modes using an ancilla reset in between preparation sequences to ensure no mode-mode entanglement. \textbf{c-f} Characteristic functions of all four prepared qunaught states and corresponding eigenvalues (bottom right) from the stabilizer operators $\hat{S}_q$ and $\hat{S}_p$ applied to the corresponding infinite-energy state. The states $\qunaught_p$ and $\qunaught_{qp}$ ($\qunaught$ and $\qunaught_{q}$) are prepared using $\hat{U}_{q_0}$ (additional $\hat{U}_{\text{shift}}$), followed by a reset of the ancilla qubit.}
    \label{fig 2: state generation}
\end{figure*}

\vspace{12pt}

\section{Experiment}
We perform experiments using two mechanical bosonic modes of a single $^{40}\text{Ca}^+$ ion confined in a room-temperature monolithic segmented Paul trap \cite{e_brucke_preparation_nodate} (Fig. \ref{fig 1: overview} a). The two modes have frequencies of $(\omega_1, \omega_2)/(2 \pi) = (2.42, 2.57)~\rm{MHz}$ and when measuring a Fock state superposition $(\ket{0}+\ket{1})/\sqrt{2}$ reliably exhibit $1/e$ coherence times longer than \SI{25}{ms}  \cite{turchette_decoherence_2000}. A resonant mode-to-mode coupling is realized by modulating the voltage applied to the DC electrodes of the trap at the difference frequency between the two modes, implementing the beamsplitter operation. This generates an effective Hamiltonian  $\hat{H}_{\rm BS} =\hbar g  (\hat q_1 \hat p_2 - \hat p_1 \hat q_2)$\cite{brown_coupled_2011, gorman_two-mode_2014, hou_coherent_2024, metzner_two-mode_2024}, which we apply for a duration $t$ such that $g t = \pi/4$ to realize a 50/50 beamsplitter operation (see Methods). We achieve a coupling rate of $g/(2\pi) =\SI{1.25}{kHz}$, which we ramp on and off adiabatically to avoid unwanted excitations, resulting in a total beamsplitter duration of \SI{400}{\micro\second}. 

For state preparation, control, and measurement of the individual mechanical modes, we utilize the electronic degrees of freedom of the ion. We define an ancillary qubit using the states $\ket{\downarrow} \equiv \ket{S_{1/2},m_j = +1/2} $ and $  \ket{\uparrow} \equiv \ket{D_{5/2},m_j = +5/2}$. The qubit states are coupled via a quadrupole transition that is driven by a narrow linewidth laser at \SI{729}{nm}. Using a bichromatic laser field resonant with the red and blue sidebands of the chosen mode of motion $j \in \{1,2\}$, we realize a state-dependent force (SDF) Hamiltonian $H_{SDF} = \hbar \eta_j\Omega\hat{\sigma}_{\phi_s}\hat{q}_{j\phi_m}/\sqrt{2}$ , with $\hat{\sigma}_{\phi_s} = \cos(\phi_s)\hat \sigma_x + \sin(\phi_s) \hat \sigma_y $, where $\hat \sigma_{x,y,z}$ are the Pauli matrices, $\hat \sigma_z = \ket{\uparrow}\bra{\uparrow} - \ket\downarrow\bra\downarrow$, $\hat{q}_{j\phi_m} = \cos(\phi_m) \hat  q_j - \sin(\phi_m) \hat p_j $ and $\eta_j$ the Lamb-Dicke parameter of the mode ($(\eta_1, \eta_2) = (0.025, 0.052)$). The phases $\phi_s$ and $\phi_m$ are set by relative phases of the two sideband tones, and allow choice of the Pauli basis and phase space direction of the displacement. For a pulse with a fixed duration, the resulting operation is  $e^{-i u \hat{\sigma}_{\phi_s}\hat{q}_{j,\phi_m}}$ with $u = \eta_j \Omega t/\sqrt{2}$, corresponding to a displacement for which the sign depends on the internal state. Circuits composed of SDFs are used for the control, preparation and measurement of bosonic states and are combined with optical pumping for ancilla reset operations and detection via state-dependent fluorescence.

\begin{figure*}[t]
    \centering
    \includegraphics[width=1\linewidth]{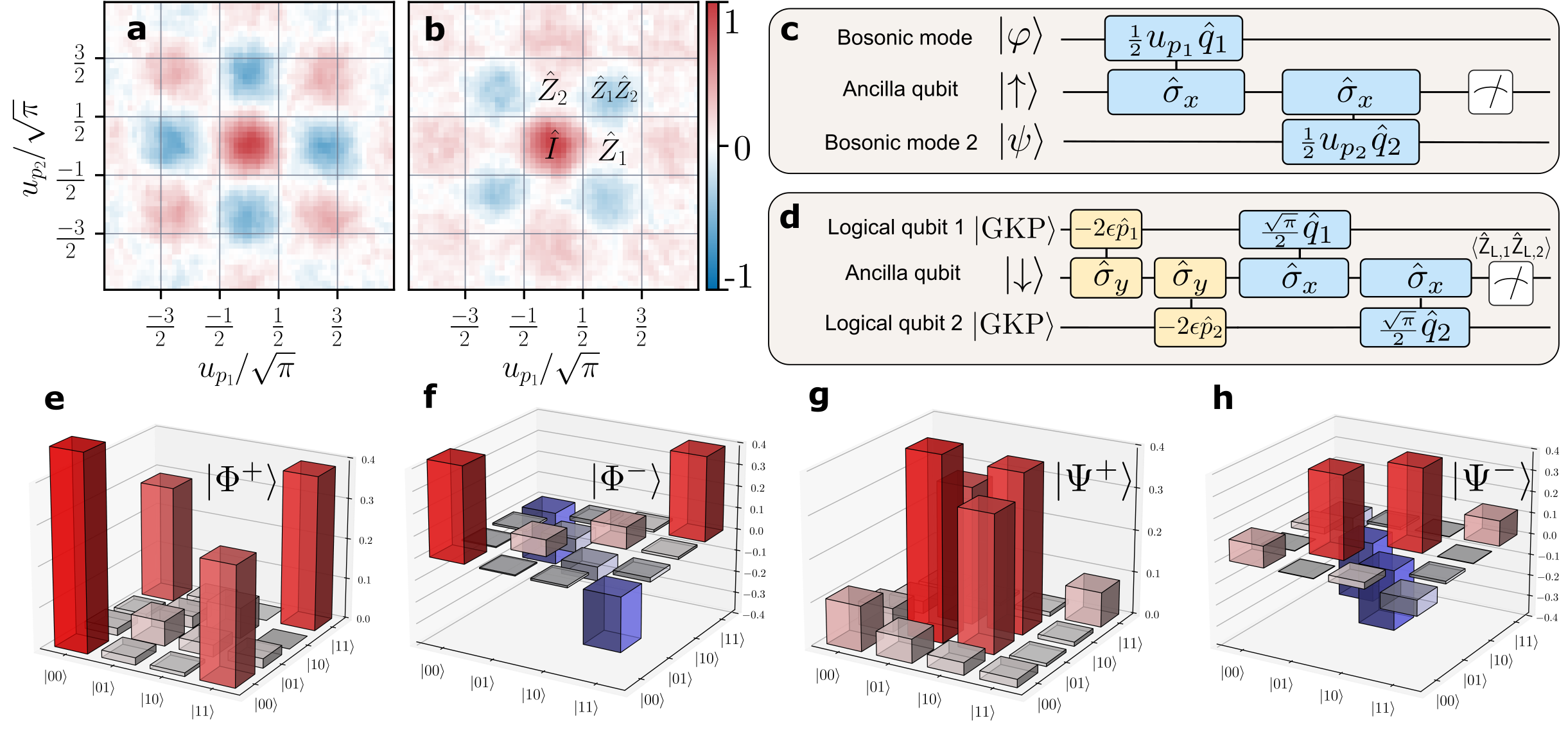}
    \caption{\textbf{Measurement of entangled GKP states.} Characteristic function of the two-mode \textbf{a} $\qunaught_{qp} \otimes \qunaught_{qp}$ state before the beamsplitter and \textbf{b} resulting Bell state after the beamsplitter operation. \textbf{c} Circuit for measuring the real parts of the two-mode characteristic function for a joint state. \textbf{d} Modified circuit for making measurements in the logical subspace including finite energy correction. We prepend conditional displacements of length $\epsilon$ (yellow) to the characteristic readout at displacement $\sqrt \pi/2$ on both modes (blue). \textbf{e-h} Reconstruction of the logical density matrices for all the different GKP Bell states with fidelities 0.72(2), 0.66(2), 0.73(2), 0.67(2) for the states $\ket{\Phi^+}, \ket{\Phi^-}, \ket{\Psi^+}, \ket{\Psi^-}$. This corresponds to an average Bell state fidelity of 0.69(1).}
    \label{fig 3: bell state characterisation}
\end{figure*}

\section{Creation of Qunaught States}
To prepare the different qunaught states, we start by squeezing each of the modes, reducing the $q$ quadrature using a parametric modulation of the trapping potential \cite{burd_quantum_2019}. Applying this modulation for \SI{75}{\micro \second} implements the unitary $\hat{S}_i(r) = e^{ir/2(\hat{q}_i\hat{p}_i+\hat{p}_i\hat{q}_i)}, i= {1,2}$, with $r\sim0.5$. We subsequently apply five SDFs sequentially (Fig.~\ref{fig 2: state generation} a) on each of the modes (Fig. \ref{fig 2: state generation} b). The sequence combines displacement lengths of $l$ and $l/2$ which define the grid structure of the state in phase space with those of $\pi/(4 l), \pi/(2 l) $ that disentangle the ancilla from the oscillator. An additional short SDF of length $\varepsilon $ accounts for the finite energy of the states \cite{hastrup_measurement-free_2021, de_neeve_error_2022}. This sequence generates $\qunaught_{q,p}$ and $\qunaught_p$. To generate $\qunaught$ and $\qunaught_q$~we use two additional SDFs of lengths $l/2$ and $\pm \varepsilon$ \cite{hastrup_measurement-free_2021}. Measurements of the characteristic functions of these states are shown in Fig. ~\ref{fig 2: state generation} c-f. To ensure that no entanglement is created between the modes during the state preparation, we reset the ancilla qubit using optical pumping between the sequences applied to each mode. The finite energy displacement $\varepsilon \ll l$ is experimentally optimized to maximize fidelity as well as the disentanglement of ancilla and motion (see Methods). We characterize single-mode states by extracting the characteristic function \cite{fluhmann_direct_2020}. The fidelity of the prepared states is limited by motional mode and ancilla dephasing. Based on simulations using characteristic decay times in our system, the loss in contrast of the interference fringes of the state is consistent with the experimental measurements including a preparation time of $436 \mu$s.

\begin{figure*}[t]
    \centering
    \includegraphics[width=1\linewidth]{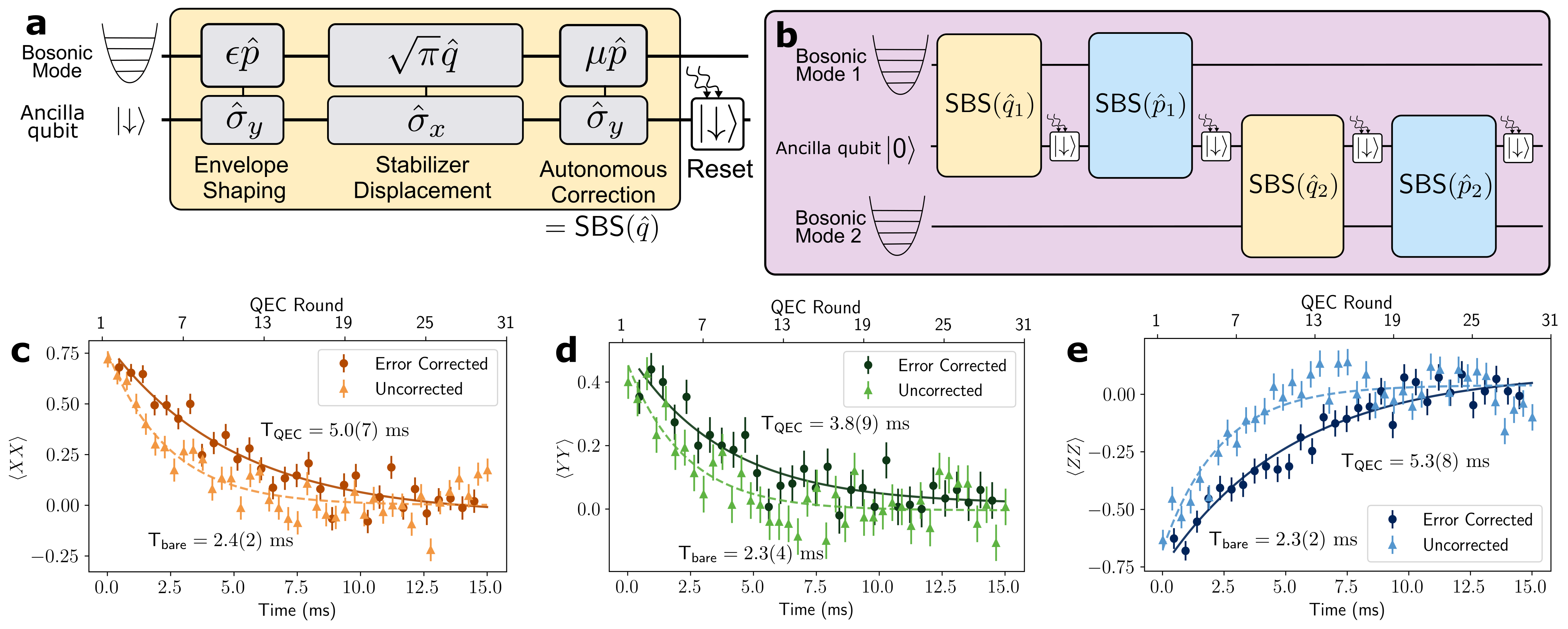}
    \caption{\textbf{Error correction of GKP Bell states.}  \textbf{a} Sequence for correcting errors on the quadrature $\hat{q}$ \cite{de_neeve_error_2022, royer_stabilization_2020}. \textbf{b} Error correction sequence consecutively correcting for errors in both  quadratures of both modes, $\hat q_1, \hat p_1, \hat q_2, \hat p_2$. \textbf{c}-\textbf{e} Lifetime measurements of the two-mode logicals for the $\ket{\Psi^+}$ GKP Bell state without/with error correction in light/dark colors. We observe a lifetime of $\SI{5.0(7)}{ms}~(\SI{2.4(2)}{ms})$, $\SI{3.8(9)}{ms}~(\SI{2.3(4)}{ms})$, $\SI{5.3(8)}{ms}~(\SI{2.3(2)}{ms})$ for the $\langle X_{L,1}X_{L,2} \rangle,\langle Y_{L,1}Y_{L,2} \rangle,\langle Z_{L,1}Z_{L,2} \rangle$ Paulis with (without) error correction giving an extension of the Bell state lifetime by a factor of 2.0(2) on average. }
    \label{fig 4: bell state qec}
\end{figure*}

\section{Bell State Generation}
Following the preparation of the particular qunaught product state, we apply the beamsplitter to generate the GKP Bell state (Fig. \ref{fig 3: bell state characterisation} a-b). To characterize the resulting entangled state, we extract selected planes of the two-mode characteristic function. This is performed by initializing the ancilla in $\ket \downarrow$ and subsequently implementing two successive SDF pulses (Fig. \ref{fig 3: bell state characterisation} c), one on each of the bosonic modes, before performing a projective measurement of the ancilla in the $\hat \sigma_z$ or $\hat \sigma_y$ basis \cite{fluhmann_direct_2020, valahu_direct_2023}. By restricting two of the four coordinates $u_{q_i}, u_{p_i}$ to 0, we characterize the initial as well as the resulting state in the $u_{p_1},u_{p_2}$ plane (Fig. \ref{fig 3: bell state characterisation} a-b).

The GKP logical and stabilizer values correspond to elements of the characteristic function at defined values of $\vec u_1, \vec u_2$. Due to the finite energy of the GKP states, a loss of contrast due to incomplete state overlap is incurred when reading out these locations using a single SDF for each mode \cite{de_neeve_error_2022, hastrup_improved_2021}. To read out logical and stabilizer values, we therefore extend methods used in finite-energy compensated readout for single modes to the two-mode case. Similar to the finite energy correction in the state initialization, this involves prepending a position-dependent rotation, using another SDF, of the ancilla qubit with a displacement  $\epsilon \ll l$ (Fig. \ref{fig 3: bell state characterisation} d, see Methods). This technique can theoretically increase the readout fidelity by up to 10$\%$ over uncompensated readout for the energy of the states considered in this work. Experimentally, decoherence of the ancilla qubit limits the increase to around 6$\%$ over uncompensated readout. With the finite-energy two-mode Pauli logicals  $\hat P_1 \hat P_2$ where $\hat P_i \in \{\hat X_{L,i}, \hat Y_{L,i}, \hat Z_{L,i}\}$ we are able to reconstruct the full two mode density matrix (see Methods) of the logical subspace. Different methods including maximum likelihood estimation and convex optimization ~\cite{riofrio_experimental_2017} return states with corresponding Bell-state fidelities of $F = 0.72(2), 0.66(2), 0.73(2)$ and $0.67(2)$ for the four states (Fig \ref{fig 3: bell state characterisation} e-h).

\section{Quantum Error Correction of the Bell state}
After preparing the entangled state, we perform multiple rounds of error correction on each GKP qubit, using the protocols previously demonstrated for a single mode \cite{royer_stabilization_2020, de_neeve_error_2022, sivak_real-time_2023} (Fig. \ref{fig 4: bell state qec} a-b). The sequence encodes stabilizer information from the oscillator into the ancilla using finite energy corrected stabilizer measurements, followed by a coherent conditional feedback correction displacement from the ancilla to the oscillator using an SDF of displacement $\mu$. This block of SDF pulses is followed by a reset of the ancilla using optical pumping. A full round of quantum error correction (QEC). shown in Fig.~\ref{fig 4: bell state qec} b, consists of sequential correction of both quadratures of a single mode followed by correction of the other (other orders of these error correction blocks were investigated, with no observable difference). One round of QEC takes $\approx$~\SI{500}{\micro \second}. The ancilla reset induces unwanted recoil due to scattered photons. To minimize this effect, we align the reset laser perpendicular to the mechanical modes used to encode information. Residual recoil from emission is small enough to be correctable by the code~\cite{de_neeve_error_2022}. 

We measure the lifetime of the expectation value of two-qubit Pauli operators $\langle X_{L,1}X_{L,2} \rangle,\langle Y_{L,1}Y_{L,2} \rangle,\langle Z_{L,1}Z_{L,2} \rangle$ to be $(5.0(7),3.8(9), 5.3(8))$~ms when undergoing error correction, and $(2.4(2), 2.3(4), 2.3(2))$~ms without error correction for the $\ket{\Psi^+}$ Bell state. This represents an increase by factors $(2.1(3), 1.7(5),2.3(4))$ respectively. The extension is least good for $\langle Y_{L,1} Y_{L,2}\rangle$, which is due to a combination of finite energy effects and an increased susceptibility of this operator to dephasing noise. While currently the performance of single-mode error correction is in a regime close to break-even with a Fock encoding of the motional mode, when correcting two modes there is twice as much recoil from repumping events, and the rate of error correction per mode is halved. Consequently, the lifetime of separable states undergoing parallel two-mode QEC is also reduced compared to the case of single-mode QEC (see Supplementary Material). 

\section{Discussion and outlook}

The beamsplitter interaction acting on GKP states, combined with error correction of the resulting entangled states, represent the primitives for fault-tolerant quantum computing with bosonic codes. The Bell state fidelities that we can achieve are primarily limited by dephasing of the motional oscillator and the ancilla qubit, which could be mitigated by improving the trap voltage stability and the magnetic field stability respectively. The duration of the beamsplitter interaction also limits the currently achievable Bell state fidelity. However, this is not a fundamental constraint: shorter operation times are attainable with stronger driving fields and longer coherence times. In our setup, the maximally attainable beamsplitter coupling is currently limited by the amplifiers in the DC voltage chain. 

Combined with single-mode squeezing and displacements, which are available in our system, the beamsplitter completes the Gaussian toolbox on the two-mode phase space~\cite{arvind_real_1995, braunstein_squeezing_2005, baragiola_all-gaussian_2019}. Realization of a fault-tolerant two-qubit gate requires applying additional squeezing operations within the gate sequence~\cite{rojkov_two-qubit_2024}. The increased noise sensitivity of squeezed GKP states reduces coherence times, posing a challenge for high-fidelity two-qubit gates.

Looking beyond codes based on a single oscillator, the dimension of the lattice of the code can be extended by incorporating additional oscillators. In trapped-ion systems, scaling up could be achieved by increasing the number of ions, with coupling between physically separate modes~\cite{brown_coupled_2011, harlander_trapped-ion_2011} or by using eigenmodes of longer ion chains~\cite{gorman_two-mode_2014, hou_coherent_2024}. 

Several multi-mode codes exist~\cite{royer_encoding_2022}. The product state $\qunaught\otimes\qunaught$, demonstrated here, is a codeword of the $D_4$ multi-mode GKP 
encoding~\cite{royer_encoding_2022} for which the 50/50 beamsplitter is a Hadamard gate up to single mode phases. Such multimode architectures would enable codes like the Tesseract code~\cite{royer_encoding_2022}, which can suppress the propagation of ancilla decay errors onto the logical subspace, a limitation that the single-mode error correction demonstrated here does not address. Alternatively, concatenation with discrete-variable codes such as the surface code could relax the required GKP squeezing levels and introduce a fault-tolerance threshold~\cite{fukui_high-threshold_2018, vuillot_quantum_2019, noh_low-overhead_2022}. 

The elements demonstrated here are also central to a range of theoretical proposals including measurement-based quantum error correction~\cite{baragiola_all-gaussian_2019, walshe_continuous-variable_2020}, teleportation-based GKP error correction~\cite{glancy_error_2006,walshe_continuous-variable_2020}, and transduction protocols that exploit the structure of GKP encodings~\cite{wang_passive_2025}.

\begin{acknowledgments}
We mourn the loss of Martin Wagener, who started building the experimental setup used in this project and greatly contributed to the implementation of the beamsplitter operation. The authors thank Brennan de Neeve for useful discussions about the QEC protocol, Hendrik Timme for contributions to the characterization of the experimental setup, Ting Rei Tan, Matthew Stafford and Xanda Kolesnikow for discussions on GKP measurement techniques, Fabian Schmid and Alfredo Ricci Vásquez for help with the laser system. This work was supported by the Swiss National Science Foundation (SNF/SNSF) under grant number 200021-22799. S.W. acknowledges financial support via the SNSF Swiss Postdoctoral Fellowship (Project no. TMPFP2$\_$210584). S.W. additionally acknowledges support from the Center for Integrated Quantum Science and Technology (IQST) and financial support from the German Research Foundation through the Emmy Noether Grant No. WE 7554/1-1, and the Carl-Zeiss-Stiftung Center for Quantum Photonics (QPhoton).
\end{acknowledgments}
\section*{Conflicts of interest}
J. H. and D.K. are advisors and shareholders of ZuriQ AG. J.H. and M. S. are shareholders of Qendra AG. The other authors declare no competing interests.

\onecolumngrid
\section{methods}
\subsection{Beamsplitter}
In order to understand how a logical Bell state is generated using qunaught states, it is useful to consider the wavefunction of the ideal states. The wavefunction of the qunaught can be written in the position basis as
\begin{equation}
    \ket{\varnothing} = \sum_{k}{ \ket{q = k \sqrt{2\pi}}}.
\end{equation}
We also consider displaced versions of this state along both $q$ and $p$. The action of these displacement operators onto a $q$ eigenstate is
\begin{equation}
    e^{-i \alpha \hat p}\ket{q} = \ket{q+\alpha} 
\end{equation}
and \begin{equation}
    e^{i \alpha \hat{q}}\ket{q}= e^{i\alpha q}\ket{q}.
\end{equation}
The result of displacing the quanught state in both position and momentum by half of the grid spacing is 
\begin{equation}
    \qunaught_{qp} = e^{i\sqrt{2 \pi}\hat{q}/2}e^{-i\sqrt{2 \pi}\hat{p}/2}\ket{\varnothing} = \sum_k{(-1)^k\ket{q = \sqrt{2\pi}(k+1/2)}}
\end{equation}
The unitary $\hat{B}$ (main text) transforms the quadratures as \cite{wang_passive_2025}
\begin{equation}
    \hat{B} |q_1\rangle |q_2\rangle = \ket{\frac{q_1 - q_2}{\sqrt{2}}} \ket{\frac{q_1 + q_2}{\sqrt{2}}}
\end{equation}
and two qunaught states are transformed as
\begin{equation}
    \hat{B} \ket{\varnothing}\ket{\varnothing} = \sum_{k,l} \ket{\sqrt{\pi}(k-l)}\ket{\sqrt{\pi}(k+l)}.
\end{equation}
For every combination of $k$ and $l$, $(k-l)$ and $(k+l)$ have the same parity, in this way the resultant state can be separated into even and odd parity which gives
\begin{equation}
    \hat{B} \ket{\varnothing}\ket{\varnothing} = \frac{1}{\sqrt 2}\left(\ket{0,0}_L+\ket{1,1}_L\right),
\end{equation}
in the GKP basis. Displacing the qunaught state gives three other possible outputs for identical input states
\begin{equation}
    \hat{B} \qunaught_{qp}\qunaught_{qp} = \sum_{k,l} (-1)^{k+l}\ket{\sqrt{\pi}(k-l)}\ket{\sqrt{\pi}(k+l+1)} = \ket{0,1}_L-\ket{1,0}_L,
\end{equation}
\begin{equation}
    \hat{B} \qunaught_{p}\qunaught_{p} = \sum_{k,l} (-1)^{k+l}\ket{\sqrt{\pi}(k-l)}\ket{\sqrt{\pi}(k+l)}= \ket{0,0}_L-\ket{1,1}_L,
\end{equation}
\begin{equation}
    \hat{B} \qunaught_{q}\qunaught_{q} = \sum_{k,l} \ket{\sqrt{\pi}(k-l)}\ket{\sqrt{\pi}(k+l+1)} = \ket{0,1}_L+\ket{1,0}_L.
\end{equation}
The result of the displacement causes the two states to have either opposite parity or to generate an opposite sign of the two parities. Each one of these outcomes represent a different logical Bell state.

Experimentally, we realize the beamsplitter by parametrically coupling the two modes. Let mode $1$ be along $x$ and mode $2$ along $y$, the Hamiltonian confining the ions in this plane may then be written as 
$$
H = \frac{p_x^2 + p_y^2}{2m} + \frac{1}{2}m \omega_x^2 x^2 + \frac{1}{2}m \omega_y^2 y^2 
$$
The coupling is then realized by applying a tilting potential $V_C = V(t) xy$ where $V(t)$ can be chosen at the frequency difference to drive the beamsplitter \cite{gorman_two-mode_2014, hou_coherent_2024}.

\subsection{Two-mode finite-energy corrected readout}
\label{sec:fe_readout}

In the GKP code, the logical information is encoded in expectation values of displacement operators, where for the ideal GKP code $\hat{Z}  = e^{i \sqrt \pi \hat q}$. The expectation value of the Paulis therefore can be related to the characteristic function \cite{fluhmann_direct_2020}, which can be directly mapped into an an ancilla by applying the unitary $\hat U_Z = \exp(i \sqrt \pi/2 \hat q \sigma_x )$ such that after application $\hat U_Z$ one has 
\begin{equation}
\langle\hat{\sigma}_z\rangle = \langle \cos(\sqrt \pi \hat q)\rangle 
\end{equation}
which reads out the logical $\hat Z$ content of the GKP state. In the case of finite-energy states, the overlap with the state and its displaced self incurs a loss of contrast proportional the envelope \cite{fluhmann_encoding_2019, hastrup_improved_2021, de_neeve_error_2022}. In order to compensate this effect, a biasing envelope pulse is added such that $\hat{U}_Z = e^{i \frac{\sqrt \pi}{2} \hat{q} \hat{\sigma}_x}e^{-i \epsilon \hat{p} \hat{\sigma}_y} $ \cite{de_neeve_error_2022}.
The action of the biasing pulse for a logical measurement is dependent on the state of the ancilla; this necessitates that the envelope pulse is applied before the measurement pulse, otherwise the envelope reduces the readout fidelity for the measurement of $\ket{1}_L$. The required duration of the biasing pulse is dependent on the envelope of the state, where the final outcome for a GKP $Z$ eigenstate is
\begin{equation}
    \langle \hat Z_L \rangle \approx \langle \hat \sigma_z \rangle = (-1)^{z} e^{-\pi \kappa^2/4}\bigg[e^{-\epsilon^2/\kappa^2}+\sin{(\epsilon \sqrt{\pi})}\bigg],
\end{equation} 
where $\kappa = e^{-r}$, with a squeeze parameter $r$ and $z \in \{0,1\}$ is the qubit state.
With a single ancilla, this technique can be extended to measurement of more than one mode, where measurements of operators like $ \hat{Z}_1\hat{Z}_2$ is required. To correct for finite energy effects from both modes two biasing pulses are needed, both of which must happen before the logical pulse giving
\begin{equation}
    \hat U_{\hat Z_{L,1} Z_{L,2}} = e^{i \frac{\sqrt \pi}{2} \hat{q}_1 \hat{\sigma}_x} e^{i \frac{\sqrt \pi}{2} \hat{q}_2 \hat{\sigma}_x} e^{-i \epsilon \hat{p}_1 \hat{\sigma}_y} e^{-i \epsilon \hat{p}_2 \hat{\sigma}_y} .
\end{equation}
The application of $\hat U_{\hat Z_{L,1} Z_{L,2}}$  biases the ancilla in a way that corrects for overlap loss in the two-mode case and maps the two-mode logical operator into the ancilla. The measurement result for $\langle Z_1Z_2 \rangle$ results in a similar increase in readout fidelity, where the result is proportional to $\langle Z_1 \rangle^2$ but $\epsilon \rightarrow 2\epsilon$ resulting from a rescaling of the phase space for the collective quadratures, giving
\begin{equation}
    \langle \hat Z_{L,1} \hat Z_{L,2} \rangle \approx \langle \sigma_z \rangle = (-1)^{z_1+z_2} e^{-\pi \kappa^2/2}\bigg[e^{-2\epsilon^2/\kappa^2}+\sin{(2\epsilon \sqrt{\pi})}\bigg].
\end{equation} 
For our GKP squeezing of $\kappa = 0.37$ (see Supplementary Material \ref{sec:gkp_prep} for definition), this correction cannot exceed 0.989 for $\langle\hat{Z}_L\hat{Z}_L \rangle$ and 0.674 for $\langle \hat{S}_z\hat{S}_z\rangle$.

We note that there are other techniques that could better approximate a logical measurement, and be more robust to errors. One such method is called the stabilizer sub-system decomposition (SSSD) readout \cite{shaw_stabilizer_2024, shaw_logical_2024}, which has been experimentally implemented\cite{matsos_universal_2025}.
This method gives the logical Pauli expectation values as a weighted sum of expectation values from various multiples of the logical operators following equations shown in ~\cite{shaw_logical_2024}. The two-mode logical expectation values are defined as the tensor products of the corresponding single-mode logical operators.  

Experimentally, unlike a single shot finite-energy measurement, this involves measuring many points in the characteristic function with both positive and negative displacements. It is possible to measure only half this number of points since the imaginary part of the characteristic function is zero for GKP states. The expectations values are evaluated with a finite truncation to the infinite sum which can introduce truncation errors. This truncation choice is a compromise between measurement time and readout fidelity, however this process always requires making a significant number of measurements and results in larger statistical uncertainties. 
Results for measurements of the GKP Bell state using SSSD are in statistical agreement in the Bell state fidelity as compared to our single shot finite-energy based measurements (see Extended Data). Recent theoretical results~\cite{singh_towards_2026, teerawat_private_communication} also provide a potential single-shot method to more faithfully measure the logical expectation values, however these methods suffer from long measurement times where environmental noise reduces readout fidelity.

\subsection{Logical state reconstruction}
In order to reconstruct the logical state density matrix, we perform a full quantum state tomography on each of the Bell states. This requires making 15 different measurements, which correspond to different points in the two-mode characteristic function. The logical density matrix is then given as
\begin{equation}
    \hat{\rho}_L = \dfrac{1}{4}\sum_{\hat P_{i,j} \in \{\hat I, \hat X, \hat Y, \hat Z\}} \langle \hat P_i \otimes \hat P_j \rangle \hat P_i\otimes \hat P,
\end{equation}
where $\hat P_j$ are the logical Pauli operators of the GKP code including the identity, measured using the finite-energy method previously described. The expectation values $\langle \hat P_i \otimes \hat P_j \rangle = \text{Tr}[\rho_L(\hat P_i \otimes \hat P_j)]$ are the measurement outcomes after measuring the state of the ancilla qubit. The results for all the Bell state tomographies are presented in the Extended Data Fig \ref{fig:additional_fig_1}. The reported uncertainties in the Bell state fidelities are determined using a bootstrap procedure where the density matrix is reconstructed 1,000 times sampling from a binomial distribution for each of the measurement outcomes. A fidelity, given by
\begin{equation}
    F(\hat \rho_{L},\hat \rho_{\text{Bell}}) = \bigg(\text{Tr}\sqrt{\sqrt{\hat \rho_{L}}\hat \rho_{\text{Bell}}\sqrt{\hat \rho_{L}}}\bigg)^2
\end{equation}
is computed for each reconstructed $\hat \rho_{L}$  with the uncertainty on the reported fidelity representing $1\sigma$ of the final distribution of fidelities.

\subsection{Calibrations}

The motional phases are referenced to 0 at the beginning of the experiment, and all subsequent SDF pulses are phase-referenced to that time. We find that this describes the oscillator well, with the exception of when using the beamsplitter. The beamsplitter operation causes a phase shift on each mode which is calibrated out using a squeezed state as a sensor. The phase $\theta$ of the beamsplitter is calibrated by changing the delay time of the triggered waveform, see Supplementary Material \ref{sm:bs_phase_calibration}. 

\newpage
\renewcommand{\figurename}{Extended Data Fig.}
\setcounter{figure}{0}

\begin{figure*}[h]
    \centering
    \includegraphics[width=1\linewidth]{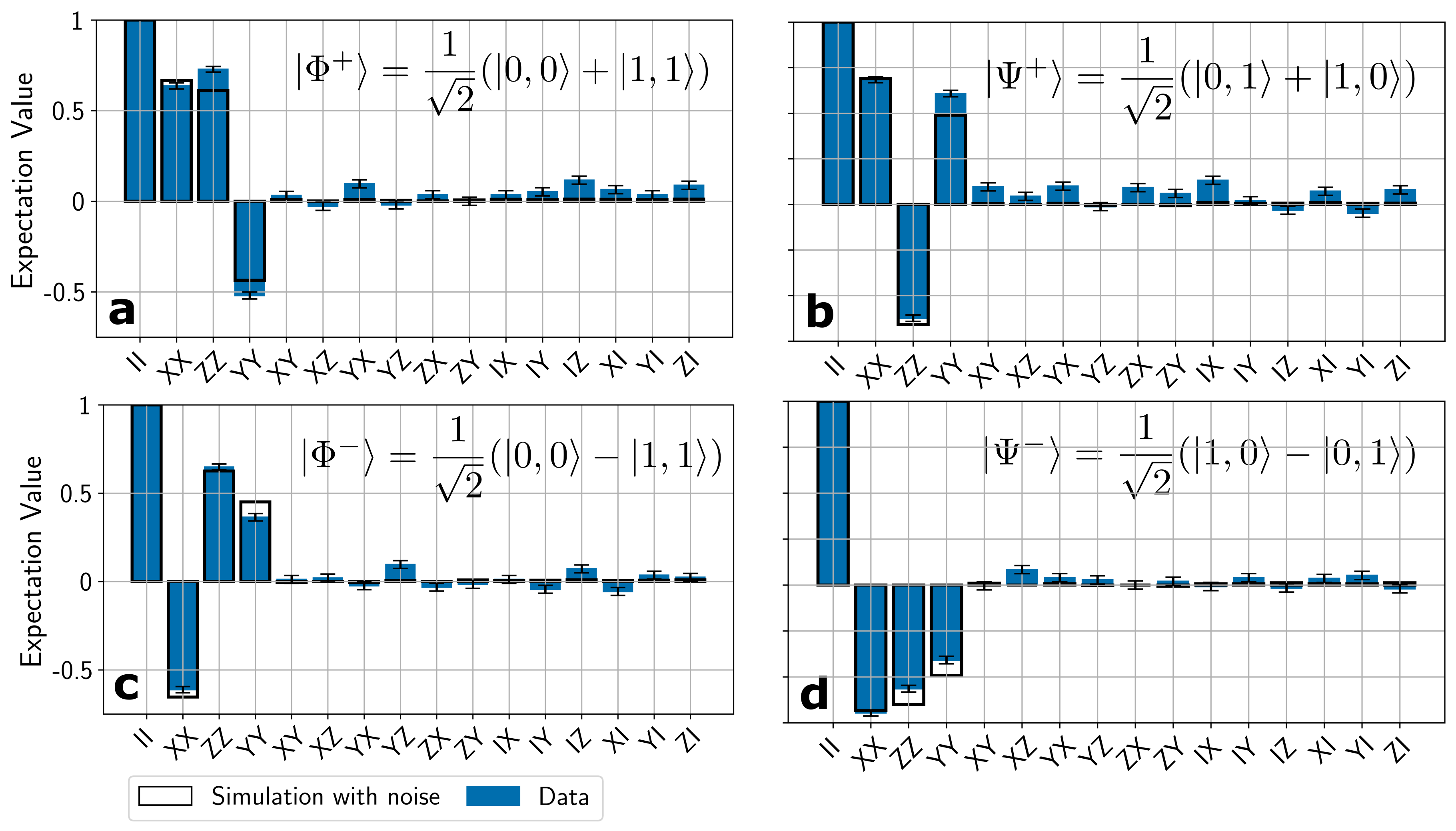}
    \caption{\textbf{a}-\textbf{d} Measured tomography data of the four Bell states described in the main text (see Fig \ref{fig 3: bell state characterisation}). The fifteen Pauli operators are measured using finite-energy readout.}
    \label{fig:additional_fig_1}
\end{figure*}

\newpage

\begin{figure}[h]
    \centering
    \includegraphics[width=0.5\linewidth]{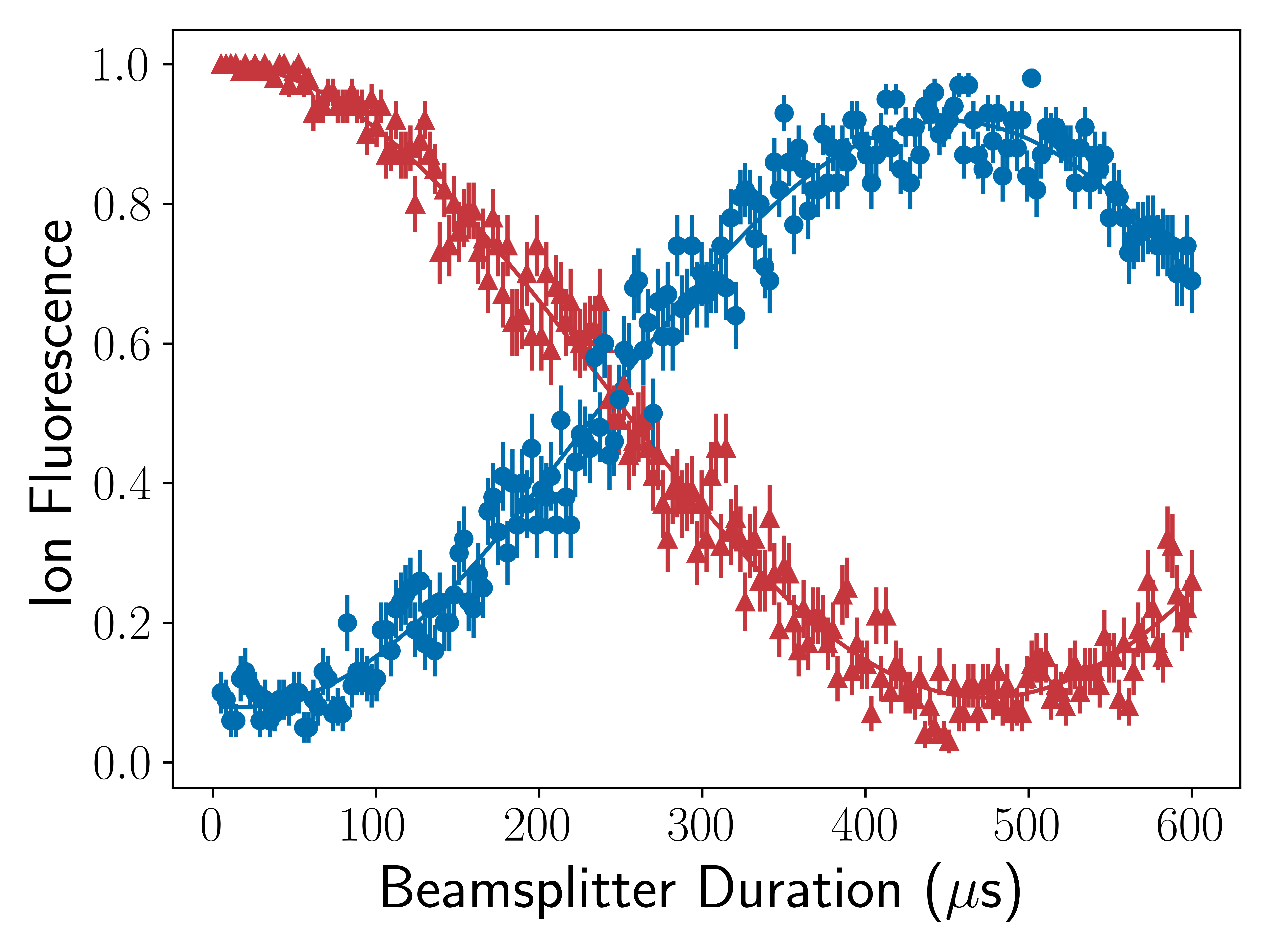}
    \caption{\textbf{Beamsplitter characterization} One of the two radial modes is prepared in the vacuum state and the other mode in a single phonon state. Upon turning on the beamsplitter interaction, the single phonon is transferred from one mode to the other. The red triangles/blue dots curves show this behavior for the phonon prepared in the radial 1/2 mode. We attribute the imperfect contrast to state preparation and readout, which is done using sequences of $\pi$ pulses on the carrier and on the red sidebands.}
    \label{fig:additional_fig_2}
\end{figure}

\newpage

\begin{figure}[H]
    \centering
    \includegraphics[width=1\linewidth]{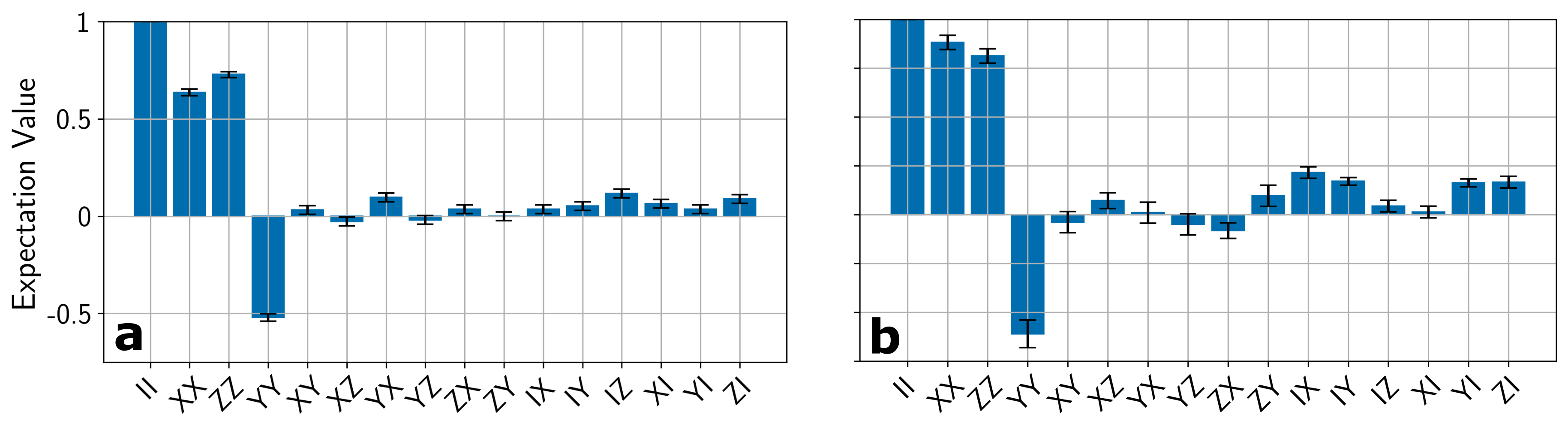}
    \caption{Quantum state tomography of the Bell state $\ket{\Phi^+}=(|00\rangle + |11\rangle)/\sqrt{2}$. Read out is performed with two-mode finite energy readout (\textbf{a}) and SSSD readout (\textbf{b}) \cite{shaw_logical_2024,matsos_universal_2025}. Fidelities are respectively 72(2)\% and 77(4)\%.}
    \label{fig:additional_fig_3}
\end{figure}

\section*{References}
\bibliography{QEC_Bell_state}

\newpage
~
\section*{Supplementary material}

\appendix

\renewcommand{\figurename}{Fig. S}
\setcounter{figure}{0}

\section{Experimental setup}
\begin{figure}[h!]
    \centering
    \includegraphics[width=0.3\linewidth]{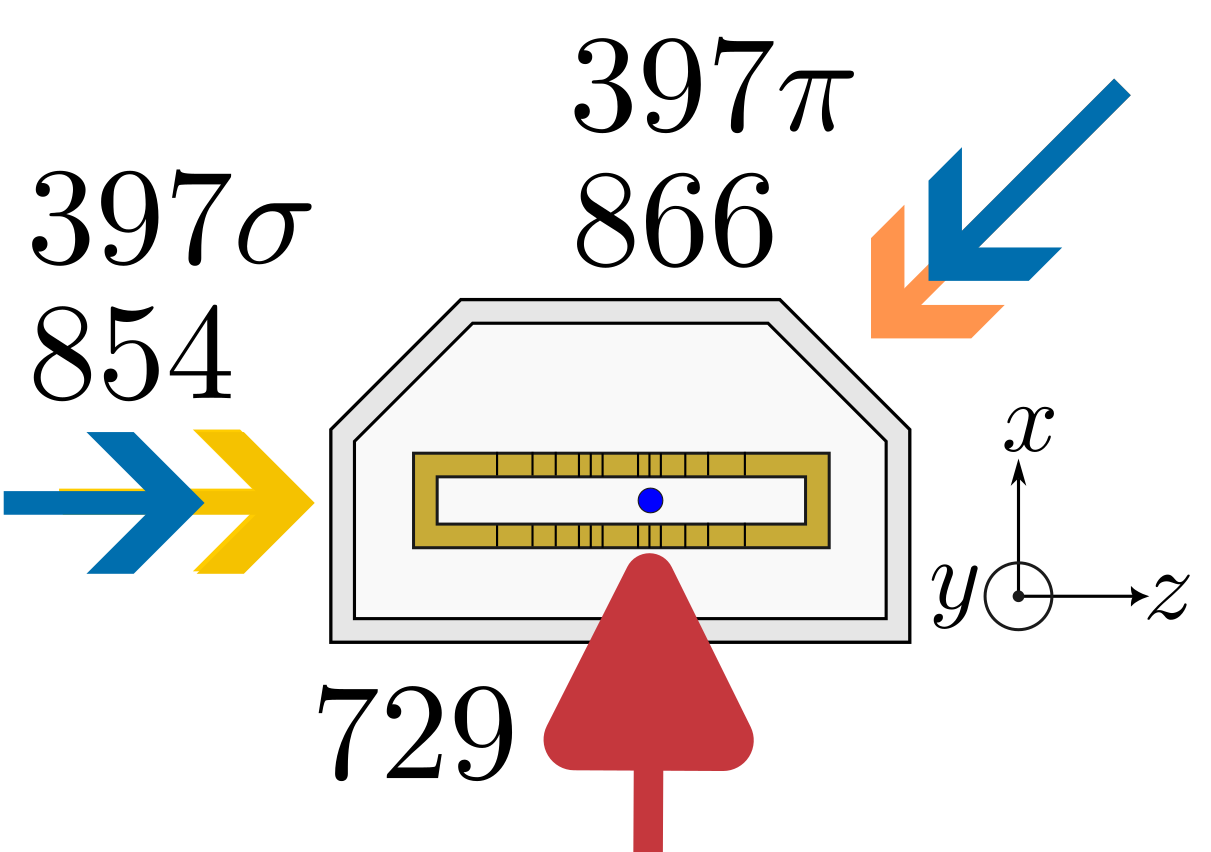}
    \caption{Top view schematic of the experimental setup. The gray box indicates the shape of the vacuum chamber. In the z direction, the ion is trapped approximately \SI{500}{\micro m} away from the trap center. The 729 nm qubit light is delivered through a re-entrant viewport and a custom lens with NA of 0.4 into a diffraction-limited spot ($\approx$\SI{1}{\micro m}). A $397\sigma$ beam is used for EIT cooling and state preparation. The $397\pi$ beam is used for detection and doppler cooling. The 854 repumper is along the axis to minimize absorbtion recoil during GKP QEC. The magnetic field is aligned along the $z$ axis.}
    \label{fig:gkp_setup}
\end{figure}
Our experiments take place in a monolithic segmented trap \cite{e_brucke_preparation_nodate}. We trap $^{40}$Ca$^+$ in a field of approximately 5 Gauss. The setup's vacuum chamber is made out of aluminium and our pressure is $\approx 10^{-10} \si{mBar}$, which we characterized by counting the number of reordering events of a co-trapped dark ion. 

We have control over 22 segments of the electrodes, four of which are on top of the structure (shims) and 18 of which form pairs of DC electrodes used to shape potentials. Our axial mode frequency is approximately \SI{598}{kHz}.

We operate with a magnetic field of approximately 5 Gauss along $z$ and with 5 laser beams (Fig. S\ref{fig:gkp_setup}). Our 729 qubit and SDF laser is tightly focused to the axis of the trap and directed using a crossed AOD setup. We characterized the fluctuations of the beam to be of less than \SI{10}{nm} standard deviation. Our maximal measured Rabi rates is of $\Omega = 5 \cdot 2\pi\si{MHz}$, in practice we operate with less power. We actively stabilize the power of the 729 nm laser. 

The 854 nm repumper is placed along the axial mode such that the absorption recoil is mainly directed along the axial mode.

\section{Spin-motion disentanglement in $\qunaught$ preparation} 
The method that we have implemented to generate $\qunaught$ states, adapted from~\cite{hastrup_measurement-free_2021}, ideally involves no projective measurements and leaves the final state of the oscillator and qubit in a product state. Experimentally, we can quantify the residual entanglement by doing a projective measurement of the ancilla qubit after the protocol. Given a symmetric state about the origin in phase space, and the action of the SDF, it is sufficient to measure only $\langle\hat \sigma_z \rangle$. In the event of perfect disentanglement we expect to find the ancilla qubit in an eigenstate of $\hat \sigma_z$. It is important that the spin basis for each applied SDF is chosen correctly. Stark shifts that arise from the SDF also must be accounted for and compensated for with a phase offset during delays between SDFs. These phases are optimized by looking at the measurement of $\langle \hat \sigma_z\rangle$ after the sequence. We report a maximum value of $\lvert \langle \hat\sigma_z \rangle \rvert > 0.8$ for the preparation of all four qunaught states, which is 5-10\% worse than results from simulations.

\section{Beamsplitter phase calibration}
\label{sm:bs_phase_calibration}
\begin{figure}[H]
    \centering
    \includegraphics[width=0.5\linewidth]{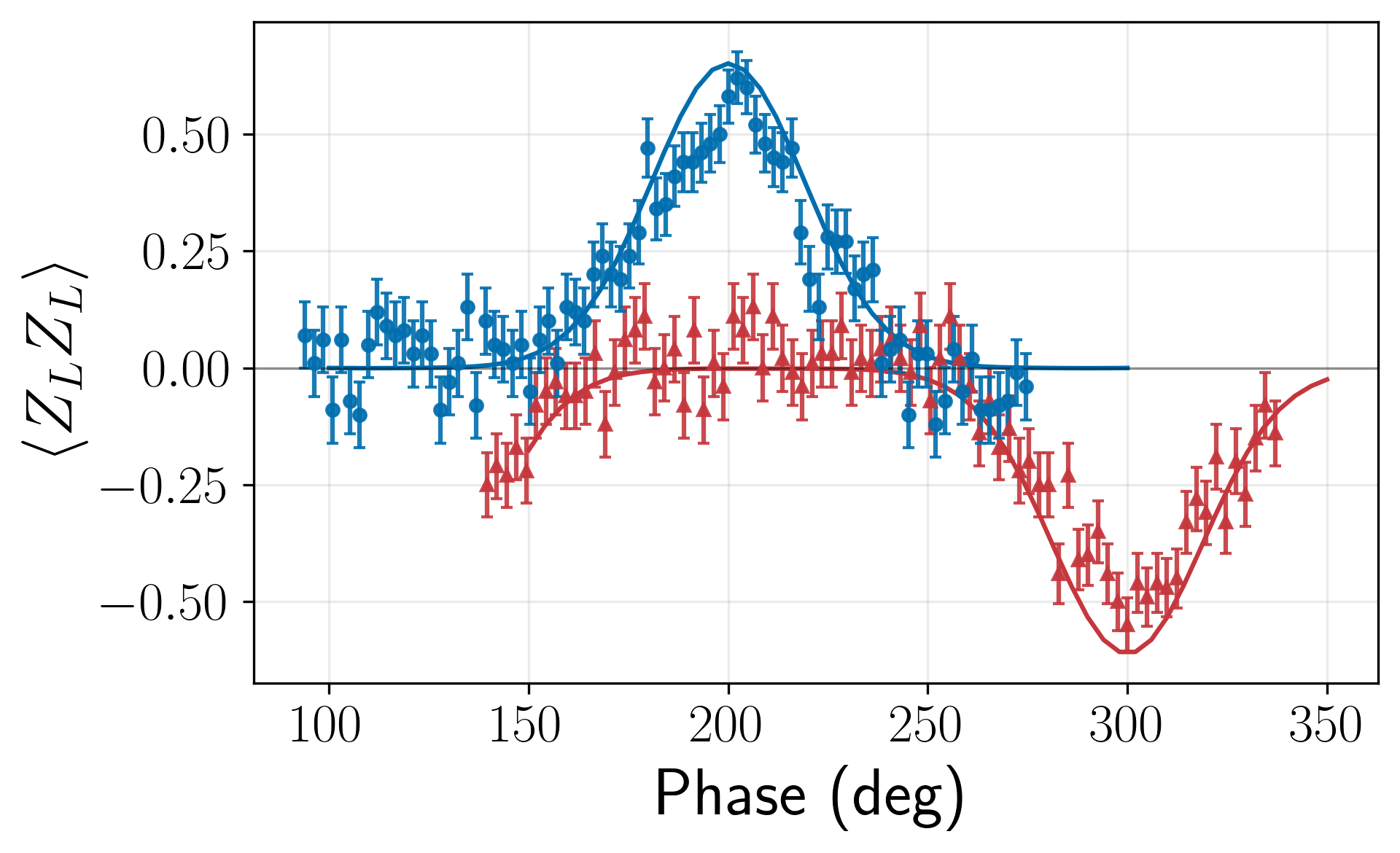}
    \caption{Calibration scans of the beamsplitter phase, determined by a delay before the beam splitter. Red triangles (blue points) are measurements of the joint logical for $\ket{\Phi^+}(\ket{\Psi^-})$ with simulation results shown as the solid red (blue) curves. }
    \label{sm:bs_phase}
\end{figure}

The beam splitter is implemented using a waveform generated from a Fastino from Creotech running custom Firmware and triggered from the experimental control system. This technique makes a direct scan of the phase of the waveform costly, as the waveform has to be re-calculated and uploaded to the Fastino for every data point. It is more efficient to instead delay the beginning of the waveform to change the beam splitter phase by $\phi_{BS} = t_{delay}(\omega_1-\omega_2)$ with mode frequencies $\omega_1,\omega_2$. Any change in the experimental control length or frequency in the oscillators means this phase has to be calibrated. Therefore, we calibrated the phase directly on the Bell state logical Pauli operator measurement. Figure ~\ref{sm:bs_phase} shows the results for these measurements for two different Bell states where the optimal phase comes at a different phase due to different state preparation lengths and different calibrated mode frequencies.

\section{GKP state preparation}
\label{sec:gkp_prep}

The Small-Big-Small (SBS) protocol, applied to a squeezed state, can also be used to dissipatively prepare GKP states and $\qunaught$ states. Those grid states are squeezed in both directions of the phase space, hence we use an effective squeezing parameter (\cite{duivenvoorden_single-mode_2017, fluhmann_encoding_2019}) to quantify the energy of the state:
\begin{equation}
    \Delta_j = \sqrt{\frac{1}{\pi} \ln\left(\frac{1}{\langle \hat{S}_j\rangle^2}\right)}
\end{equation}
where $\hat{S}_j$ is the finite-energy corrected measurement of the stabilizer in the quadrature j. 
Preparing the state with the SBS protocol yields effective squeezing results comparable with the deterministic method discussed in the main text, for $\qunaught$ and GKP X and Z eigenstates. Specifically, an effective squeezing of $\Delta \sim 0.55$ ($\sim 8$ dB calculated as $10\log_{10} {(2/\Delta_j^2)}$) is achieved for single-mode $\qunaught$, and $\Delta \sim 0.37$ ($\sim 11.5$ dB) for single-mode GKP X and Z eigenstates.
For the GKP Y eigenstate, a higher contrast is reached by preparing a $\qunaught_q$, and virtually updating the oscillator's phase by 45°. 
\section{Squeezing}

Squeezing can be realized by parametrically modulating the confining potential of the oscillator at twice its rest frequency, i.e. 

$$H = \frac{\hat P^2}{2m} + \frac{k}{2}\hat Q^2 = \frac{\hat P^2}{2m} + \frac{1}{2}\left(k_0 + k_1 \cos\left(2\omega_0t + \phi\right)\right)\hat Q^2$$
with $\omega_0 = \sqrt{k_0/m}$. Using the raising and lowering operators 
$$\hat Q = \left(\frac{\hbar}{2m\omega_0}\right)^{1/2}(\hat a^\dagger + \hat a), \qquad \hat P = i\left(\frac{\hbar \omega m}{2}\right)^{1/2}(\hat a^\dagger - \hat a)$$
we can rewrite the Hamiltonian as 
$$
H = \left[\hbar\omega_0\left(1 + \frac{k_1}{k_0}\cos(2\omega_0 t + \phi)\right)\right]\left(\hat a^\dagger \hat a + \frac{1}{2}\right) + \hbar\omega_0 \frac{k_1}{k_0 \cdot 2}\cos(2\omega_0 t + \phi)(\hat a^{\dagger 2} + \hat a^2)$$
which after going into the rotating frame of the oscillator and performing a rotating wave approximation yields the the effective squeezing Hamiltonian 

$$
	H_{sq} = \frac{\hbar g}{2}\left(\hat a^2e^{-i\phi} + (\hat a^\dagger)^2 e^{i\phi}\right)
$$
with the squeezing parameter $\quad r = g t e^{i\phi}, \quad g = \frac{\omega_0 k_1}{2k_0}$, where $t$ is the time for which the Hamiltonian was applied. 

\begin{figure}[h]
    \centering
    \includegraphics[width=0.5\linewidth]{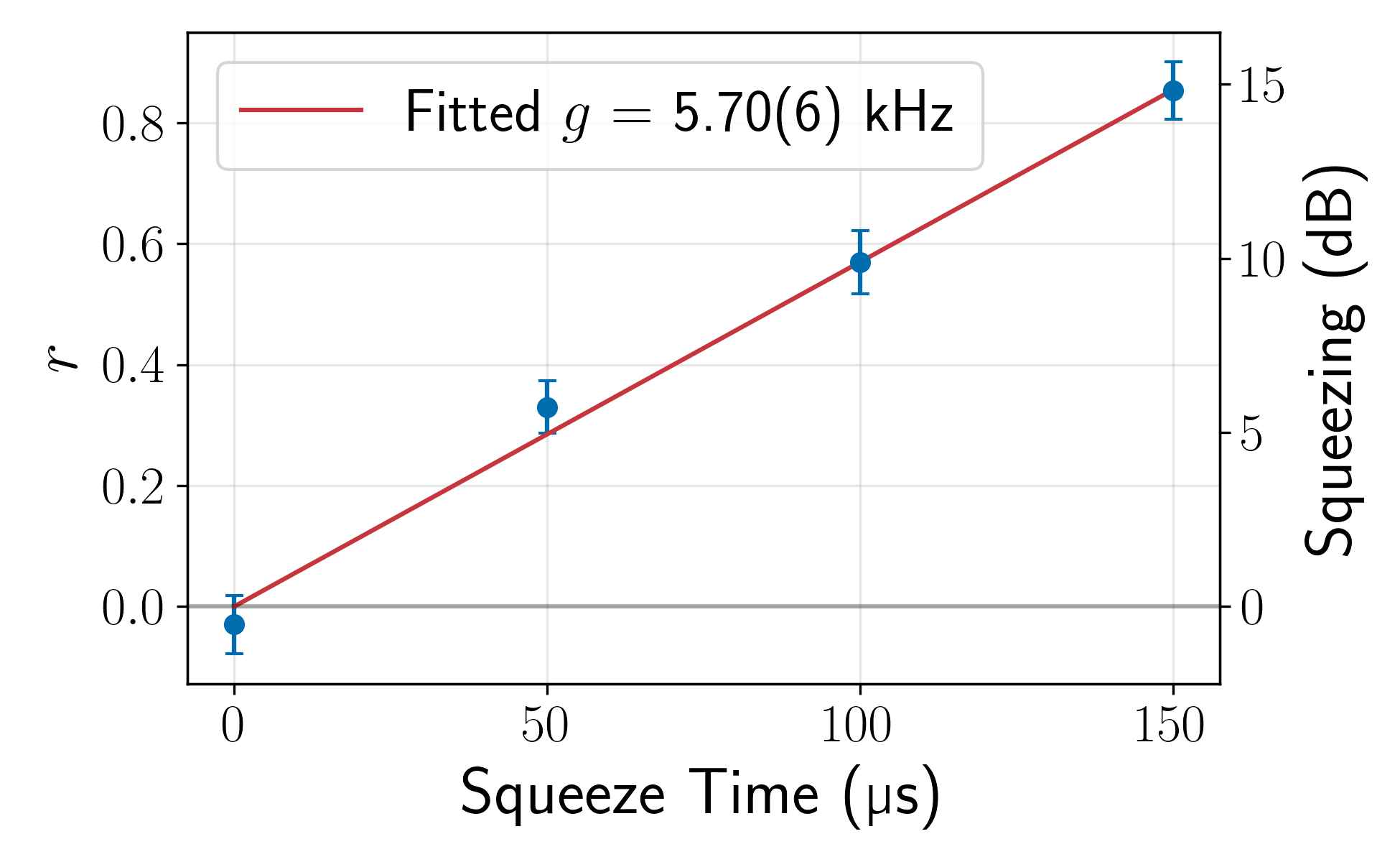}
    \caption{Characterization of the squeezing coupling constant.  The squeezing parameter $r$ (blue dots) is extracted as $ r = \ln(\sigma_l/\sigma_s)/2$, where $\sigma_s$ ($\sigma_l$) is the width of the short (long) axis of the characteristic function of the squeezed state after time $t$, extracted using characteristic function tomography. An exponential fit of the width ratio gives us the coupling strength $g = r/t$ (red line). We verify that the product of the width (as proxy for heating) remains constant.}
    \label{fig:sm_squeezing}
\end{figure}

In order to investigate the quality of our squeezing operation, we apply the squeezing Hamiltonian for a given time and then probe the long as well as the short axis of the squeezed state using a characteristic function tomography. We find that the squeezing Hamiltonian's coupling strength is $g = $ \SI{5.7\pm0.06}{kHz}, and that we can reliably produce a squeezed state with $r=0.8$, equivalent to \SI{14.8}{dB} of squeezing (Fig. S\ref{fig:sm_squeezing}). 

\section{Quantum error correction characterization}
\label{sec:qec_charac}

We evaluate the performance of quantum error-correction by comparing the lifetime of single-mode GKP states against GKP product states. 
When correcting two modes, there are several ways to sequence the SBS rounds. We tested both a mode-interleaved sequence ($\hat{S}_{q_1}, \hat{S}_{q_2}, \hat{S}_{p_1}, \hat{S}_{p_2}$) and a mode-sequential sequence ($\hat{S}_{q_1}, \hat{S}_{p_1}, \hat{S}_{q_2}, \hat{S}_{p_2}$), but observed no statistical difference, which we attribute to the quadratures being independent from each other. All data presented in this paper follows the mode-sequential QEC scheme. \\

The lifetime of the GKP states is mainly limited by the frequency drifts of the oscillators. We observe that uncorrected states exhibit coherent revivals of logical information, indicating miscalibration or coherent noise. We do not observe such revivals when performing QEC, as it pins down the logical information in the frame of the control system, and scrambles the content that leaves it. 

\begin{figure*}[h]
    \centering
    \includegraphics[width=1\linewidth]{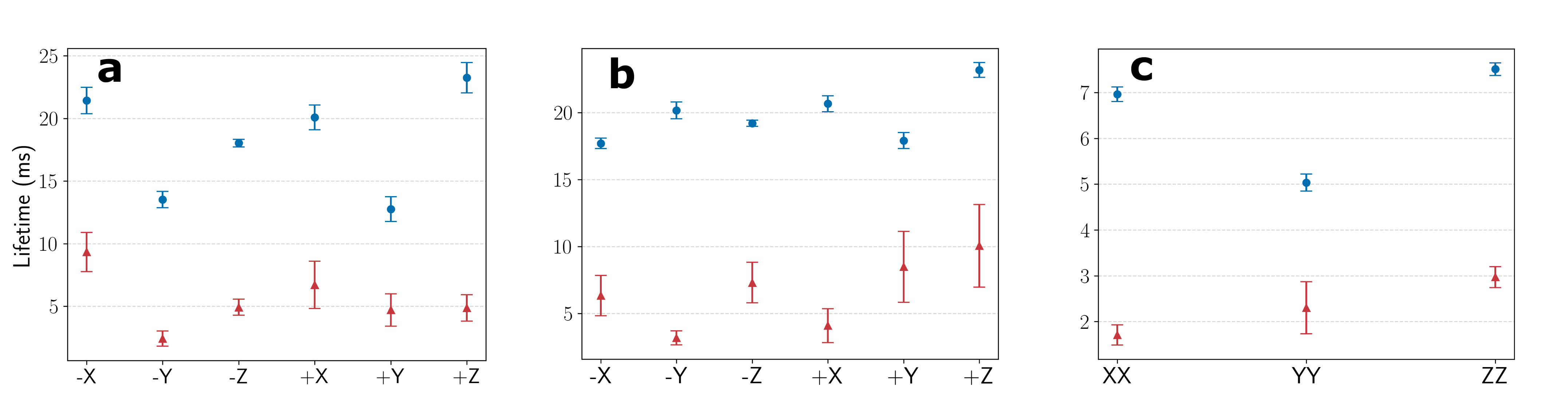}
    \caption{GKP error-corrected (blue dots) and uncorrected (red triangles) lifetimes for mode 1 (a) and mode 2 (b) measured independently. Two-mode GKP lifetimes read out with finite-energy correlated readout (c).}
    \label{fig:qec_lifetimes}
\end{figure*}

For single-mode QEC, we observe a lifetime improvement factor of 3 to 4 compared to the uncorrected case. Both modes exhibit similar performance (see Fig. \ref{fig:qec_lifetimes} a-b). The GKP lifetimes of the $\pm X$ and $\pm Z$ eigenstates are on the order of \SI{20}{ms}. The Fock state superposition $1/\sqrt2(\ket 0 + \ket 1)$ has a dephasing time that consistently exceeds \SI{25}{ms}.

When reading out correlations of a product state, the lifetime improvement reaches a factor of 2 to 3 (Fig. \ref{fig:qec_lifetimes} c), as the performance becomes limited by the simultaneous frequency tracking of both oscillators.

\end{document}